\documentclass[twocolumn,showpacs,preprintnumbers,amsmath,amssymb]{revtex4}
\usepackage{graphicx}
\usepackage{dcolumn}
\usepackage{bm}
\usepackage{amssymb}
\usepackage{amsmath}

\begin{document}

\title{Extended quantum diffusion approach to reactions of astrophysical interests}
\author{V.V.Sargsyan$^{1,2}$, G.G.Adamian$^{1}$, N.V.Antonenko$^1$, and H. Lenske$^2$}
\affiliation{$^{1}$Joint Institute for Nuclear Research, 141980 Dubna, Russia\\
$^{2}$Institut f\"ur Theoretische Physik der Justus--Liebig--Universit\"at,
D--35392 Giessen, Germany
}
\date{\today}

\begin{abstract}
The quantum diffusion approach is extended to low energy fusion (capture) reactions of light- and medium-mass nuclei.
The dependence of the friction parameter on  bombarding energy is taken into account.
A simple analytic expression  is obtained  for the capture probability at extreme sub-barrier energies.
The calculated   cross-sections are in a good agreement with the experimental data.
The fusion excitation functions calculated within the quantum diffusion
and WKB approaches are compared and presented in
the astrophysical $S$-factor  representation.
\end{abstract}


\maketitle

\section{INTRODUCTION}

Fusion reactions at energies near and below the Coulomb barrier
have been an object of extensive experimental and theoretical studies in the past decades \cite{HMOU,BackRep,CantoRep,Beck}.
Indeed, the heavy-ion fusion allows us to extend the periodic table beyond the elements that can not be
synthesized using neutrons and light charged particles. The fusion of light- and medium-mass nuclei
plays an important role in the evolution of massive stars where the behavior of fusion excitation function
at extreme sub-barrier energies determines the reaction rates.
For example, towards the end of stellar life-cycle  the elements up to the iron can be synthesized.
These reactions drives the nucleosynthesis and generates the energy in novae, supernovae, and close binary stars  \cite{IJMPour}.
In Refs. \cite{MazarakisC12C12,CujecC12O16,ChristensenC12O16,HulkeO16O16,WuO16O16,ThomasO16O16,KuronenO16O16,HighC12C12,AguileraC12C12,Desco,Gasques2007zzb,
Gasques2007zzz,AlexisC12C1200,JiangSi28Si30,Esbensen,FangC12O16,JiangC12C12,MontagnoliC12Si30,Tum,AlexisC12C12,KhoaC12C12,ZickC12C12}, the fusion
reactions involving light nuclei at low energies  were investigated both experimentally and theoretically.
The  recent developments in fusion reactions both in experiment  and theory
are presented in Refs. \cite{BackRep,CantoRep,Beck} and references therein.

For light and medium-mass nuclei the fusion is governed by the penetrability of
 colliding nuclei through the Coulomb and centrifugal barrier (so called capture).
If the collision occurs at energies permitting very large angular momentum, there is a possibility
that the formed dinuclear system decays after the capture stage.
However, at energies near and below the Coulomb barrier, the contribution of large angular momenta
to   fusion can be disregarded. Therefore, the description of   fusion of these nuclei is reduced to
the description of the capture of   projectile  by   target-nucleus.

To study the capture (fusion) process in heavy-ion reactions, the quantum diffusion approach, based on the quantum
master-equation for the reduced density matrix, has been suggested in Refs. \cite{our,EPJSub,EPJSub1}.
In this approach the collisions of nuclei are treated in terms of a single collective variable: the
relative distance $R$ between the colliding nuclei. The coupling of the relative motion to the
excitation of various channels, such as non-collective single-particle excitations, low-lying
collective modes (dynamical quadrupole and octupole excitations of the target and projectile) lead
to the fluctuation and dissipation effects. Hence, many quantum-mechanical and non-Markovian
effects, accompanying the passage through the potential barrier, are considered in our formalism.
The nuclear deformation effects are taken into account through the dependence
of the nucleus-nucleus potential on the deformations and mutual orientations of the
colliding nuclei \cite{EPJSub2,PhysPartNuc2016}.

As shown in Refs. \cite{EPJSub,EPJSub1,EPJSub2}, our model successfully describes the capture (fusion) cross section
in heavy-ion collisions at energies near and below the Coulomb barrier. In the present work we
extend our approach to describe the capture (fusion) of light and  medium-mass nuclei at energies well below
the Coulomb barrier.
Our aim is to
 calculate the fusion cross sections for
nuclei of interest and importance for stellar burning.
So, the approach is applied to low-energy fusion reactions with carbon, oxygen, and silicon nuclei.


\section{Formalism of the quantum diffusion approach}
The capture cross section is the sum of the partial capture cross sections~\cite{EPJSub,EPJSub1,EPJSub2,PhysPartNuc2016}
\begin{eqnarray}
\sigma_{\rm cap}(E_{\rm c.m.})&=&\sum_{J}^{}\sigma_{\rm cap}(E_{\rm
c.m.},J)\nonumber\\&=& \pi\lambdabar^2
\sum_{J}^{}(2J+1)\int_0^{\pi/2}d\theta_1\sin(\theta_1)\nonumber\\
&\times&\int_0^{\pi/2}d\theta_2\sin(\theta_2) P_{\rm cap}(E_{\rm
c.m.},J,\theta_1,\theta_2),
\label{1a_eq}
\end{eqnarray}
where $\lambdabar^2=\hbar^2/(2\mu E_{\rm c.m.})$ is the reduced de Broglie wavelength, $\mu=m_0A_1A_2/(A_1+A_2)$ is the reduced mass ($m_0$ is the nucleon mass),
and the summation occurs over  possible values of   angular momentum $J$ at a given bombarding energy $E_{\rm c.m.}$.
Knowing the potential of the interacting nuclei for each orientation defined by the angles $\theta_i (i=1,2)$, one can calculate the partial capture probability
$P_{\rm cap}(E_{\rm c.m.},J,\theta_1,\theta_2)$ which is the probability to penetrate throw the potential barrier in the relative coordinate $R$  at a given $J$.
$P_{\rm cap}$ is obtained by integrating the propagator $G$ from the initial state $(R_0,P_0)$ at time $t = 0$
to the final state $(R, P)$ at time $t$ ($R$ is defined with respect to the position $R_b$ of the Coulomb barrier and
$P$ is the conjugate momentum):
\begin{eqnarray}
P_{\rm cap}&=&\lim_{t\to\infty}\int_{-\infty}^{R_{\rm in}}dR\int_{-\infty}^{\infty}dP\  G(R,P,t|R_0,P_0,0)\nonumber\\
&=&\lim_{t\to\infty}\frac{1}{2} {\rm erfc}\left[\frac{-R_{\rm in}+\overline{R(t)}}
{{\sqrt{\Sigma_{RR}(t)}}}\right].
\label{1ab_eq}
\end{eqnarray}
Here, we use the propagator
\begin{eqnarray}
G=\pi^{-1}|\det {\bf \Sigma}^{-1}|^{1/2} \exp(-{\bf q}^{T}{\bf \Sigma}^{-1}{\bm q}),
\label{prop}
\end{eqnarray}
where ${\bf q}^{T}=[q_R,q_P]$, $q_R(t)=R-\overline{R(t)}$, $q_P(t)=P-\overline{P(t)}$, $\overline{R(t=0)}=R_0$,
$\overline{P(t=0)}=P_0$, $\Sigma_{kk'}(t)=2\overline{q_k(t)q_{k'}(t)}$, $\Sigma_{kk'}(t=0)=0$ and $k,k'=R,P$,
obtained in  Ref. \cite{DMDadonov} for a local inverted oscillator which replaces
the real nucleus-nucleus potential in the variable $R$. The frequency $\omega$ of this local inverted oscillator with an internal turning point
$R_{\rm in}$ is defined from the condition of   equality of the classical actions of approximated and real potential barriers of
the same height  at given $E_{\rm c.m.}$ and $J$.
Note  that this procedure leads to the frequency depending on $E_{\rm c.m.}$  and $J$.
This local replacement of the real potential by the inverted oscillator with energy-dependent frequency
is well justified for  heavy-ion reactions at energies near and below the Coulomb barrier
\cite{Hofman,VAZ,EPJSub,EPJSub1,EPJSub2,PhysPartNuc2016}.

As at $t\to\infty$ the internal turning point  $R_{\rm in}\ll\overline{R(t)}$,  the capture cross section is defined by the ratio of the mean
value of the collective coordinate $\overline{R(t)}$ and its variance $\Sigma_{RR}(t)$.
For the explicit expressions for $\overline{R(t)}$  and $\Sigma_{RR}(t)$ we refer to our previous studies in Refs.\cite{VAZ,EPJSub,EPJSub1,EPJSub2,PhysPartNuc2016}.
Using the Hamiltonian of the system, which includes the collective subsystem, the environment
(which mimics the internal excitations) and the coupling between the collective subsystem and the environment,
a system of non-Markovian Langevin equations for the collective coordinates was derived.
These equations of motion for the collective subsystem satisfy the quantum fluctuation - dissipation relations and
contain the influence of quantum, dissipative and  non-Markovian effects on the collective motion \cite{VAZ,our}.
The expressions for the $\overline{R(t)}$  and $\Sigma_{RR}(t)$ are
\begin{eqnarray}
\overline{R(t)}&=&A_tR_0+B_tP_0,\nonumber\\
\Sigma_{RR}(t)&=&\frac{4\hbar^2\tilde\lambda\epsilon\mu\gamma^2}{\pi}\int\limits_{0}^{t}d\tau^{'}B_{\tau^{'}} \int\limits_{0}^{t} d\tau^{''}B_{\tau^{''}}
\int\limits_{0}^{\infty} d\Omega \frac{\Omega}{\Omega^2+\gamma^2}\nonumber\\
&\times&\coth\left[\frac{\hbar\Omega}{2T}\right]
\cos[\Omega (\tau^{'}-\tau^{''})],\nonumber\\
B_t&=&\frac{1}{\mu}\sum_{i=1}^{3}\beta_i(s_i+\gamma)e^{s_it},\nonumber\\
\quad A_t&=&\sum_{i=1}^{3}\beta_i[s_i(s_i+\gamma)+2\hbar\tilde\lambda\epsilon\gamma]e^{s_it}.
\label{RZ_eq}
\end{eqnarray}
Here, $\Sigma_{RR}(0)=0$,  $A_0=1$, and $B_0=0$.
In Eqs.(\ref{RZ_eq}),
$\beta_1=[(s_1-s_2)(s_1-s_3)]^{-1}$,  $\beta_2=[(s_2-s_1)(s_2-s_3)]^{-1}$ and
$\beta_3=[(s_3-s_1)(s_3-s_2)]^{-1}$, and
$s_i$ are the real roots ($s_1\ge 0> s_2 \ge s_3$) of the following equation
\begin{eqnarray}
(s+\gamma)(s^2- \epsilon^2)+2\hbar\tilde\lambda\epsilon\gamma s=0.
\label{Root_eq}
\end{eqnarray}
The parameters $\gamma$, $\epsilon$ and $\tilde\lambda$  determine the characteristics of the system.
The values of $\gamma^{-1}$ is the memory time of dissipation of relative motion energy by the internal subsystem
or is the inverse bandwidth of the
internal subsystem excitations.
The non-Markovian effects appear in the calculations through $\gamma$. The instantaneous dissipation corresponds to taking $\gamma\to\infty$.
The parameter $\epsilon$  defines the initial frequency of the collective subsystem and $\tilde\lambda$
determines the average coupling strength of the collective subsystem with internal excitations.
To set these parameters \cite{VAZ,our}, we use the asymptotic values of the friction coefficient
\begin{eqnarray}
\lambda=-(s_1+s_2)
\label{Asm_fr1}
\end{eqnarray}
and potential frequency
\begin{eqnarray}
\omega=\epsilon\left(\frac{(s_1+\gamma)(s_2+\gamma)}{(s_1+\gamma)(s_2+\gamma)-2\hbar\tilde\lambda\gamma\epsilon}\right)^{1/2}.
\label{Asm_fr2}
\end{eqnarray}
Note, that $\omega$ takes into account the renormalization of the initial frequency due to
the coupling to the internal excitations.
So, in the asymptotic limit $t\rightarrow \infty$, the friction $\lambda$ and frequency $\omega$ are related to the
parameters  $\gamma$, $\epsilon$, $\tilde\lambda$, and the roots $s_{1,2}$ of Eq. (\ref{Root_eq}).
Setting the values of $\lambda$, $\omega$, and $\gamma$, we determine the dynamics of the system.
The use of  asymptotic values of $\lambda$ and $\omega$ is justified,
since the characteristic time of reaching them
is much shorter
than the characteristic time of  capture.

Equations ~(\ref{1ab_eq}), (\ref{RZ_eq}),  (\ref{Asm_fr1}), and (\ref{Asm_fr2})
lead to the analytic expression for the capture probability:
\begin{eqnarray}
P_{\rm cap}=
\frac{1}{2} {\rm erfc}\left[\left(\frac{\pi s_1(\gamma-s_1)}{2\hbar\mu(\epsilon^2-s_1^2)}\right)^{1/2}
\frac{\mu\epsilon^2 R_0/s_1+P_0}
{\left[\gamma \ln(\gamma/s_1)\right]^{1/2}}\right].
\label{PC_eq}
\end{eqnarray}
In the derivation of Eq. (\ref{PC_eq}) the limit of low temperatures ($T\to 0$) was used,
which is suitable for sub-barrier fusion.
Note, that the friction $\lambda$ and internal excitation width  $\gamma$ are related.
If the coupling with internal degrees of freedom is disregarded, $\lambda\to 0$,
then the limit $\gamma\to \infty$
results in the Markovian dynamics.
In the case of
\begin{eqnarray}
\frac{\lambda}{\omega}\ln(\gamma)\to {\rm const}
\label{glo_eq}
\end{eqnarray}
at $\lambda\to 0$,
the well-known quantum-mechanical barrier transmission probability is obtained
$$P_{\rm cap}\sim \exp[-2\pi(V_b-E_{\rm c.m.})/\hbar\omega].$$

\section{Nucleus-nucleus potential}

In the case of collision of deformed nuclei the effective nucleus-nucleus  potential reads as:
\begin{eqnarray}
V=V_N+V_C+\frac{\hbar^2 J(J+1)}{2\mu R^2},
\label{pot}
\end{eqnarray}
where $V_{N}$, $V_{C}$, and the last summand stand for  the nuclear, Coulomb, and centrifugal potentials, respectively ~\cite{poten}.
The potential depends on the relative distance $R$ between the center of mass of two interacting nuclei,
masses $A_i$, charges $Z_i$ and radii $R_i$ of the nuclei ($i=1,2$), the orientation angles $\theta_i$ of the
deformed (with the quadrupole deformation parameters $\beta_2^{(i)}$) nuclei and angular momentum $J$.
For deformed nuclei, the static quadrupole deformation parameters  are taken from Ref.~\cite{Ram}.
For the nuclear part of  potential,
\begin{eqnarray}
V_N=\int\rho_1(\bold
{r_1})\rho_2(\bold{R}-\bold{r_2})F(\bold{r_1}-\bold{r_2})d\bold{r_1}d\bold{r_2},
\end{eqnarray}
the double-folding formalism is used,
where
$F(\bold {r_1}-\bold{r_2})=C_0[F_{\rm in}\frac{\rho_0(\bold{r_1})}{\rho_{00}}+F_{\rm
ex}(1-\frac{\rho_0(\bold{r_1})}{\rho_{00}})]\delta(\bold{r_1}-\bold{r_2})$
is the density-depending effective nucleon-nucleon interaction and
$\rho_0(\bold{r})=\rho_1(\bold{r})+\rho_2(\bold{R}-\bold{r})$,
$F_{\rm in,ex}=f_{\rm in,ex}+f_{\rm in,ex}^{'}\frac{(N_1-Z_1)(N_2-Z_2)}{(N_1+Z_1)(N_2+Z_2)}$. Here,
$\rho_i(\bold{r_i})$  and $N_i$  are the nucleon
densities and neutron numbers of the light and the heavy
nuclei of the dinuclear system.
Our calculations are performed with the following set of parameters: $C_0=$
300 MeV fm$^3$, $f_{\rm in}=$ 0.09, $f_{\rm ex}=$ -2.59,
$f_{\rm in}^{'}=$ 0.42, $f_{\rm ex}^{'}=$ 0.54 and $\rho_{00}=$
0.17 fm$^{-3}$~\cite{poten}.
The densities of the nuclei are taken in the two-parameter symmetrized Woods-Saxon form
with the nuclear radius parameter $r_0$=1--1.15 fm and
the   diffuseness parameter $a$=0.47--0.56 fm depending on the charge
and mass numbers of the nucleus~\cite{poten}.

The Coulomb  interaction of two quadrupole deformed nuclei reads as
\begin{eqnarray}
&&V_{C}=\frac{Z_1Z_2e^2}{R}\nonumber\\
&+&\left(\frac{9}{20\pi}\right)^{1/2}\frac{Z_1Z_2e^2}{R^3}\sum_{i=1,2}R_i^2\beta_2^{(i)}
\left[1+\frac{2}{7}\left(\frac{5}{\pi}\right)^{1/2}\beta_2^{(i)}\right]\nonumber\\
&\times&P_2(\cos\theta_i),
\label{32ab_eq}
\end{eqnarray}
where $P_2(\cos\theta_i)$ is the Legendre polynomial.

\begin{figure}[h]
\centering
\includegraphics[angle=0, width=1\columnwidth]{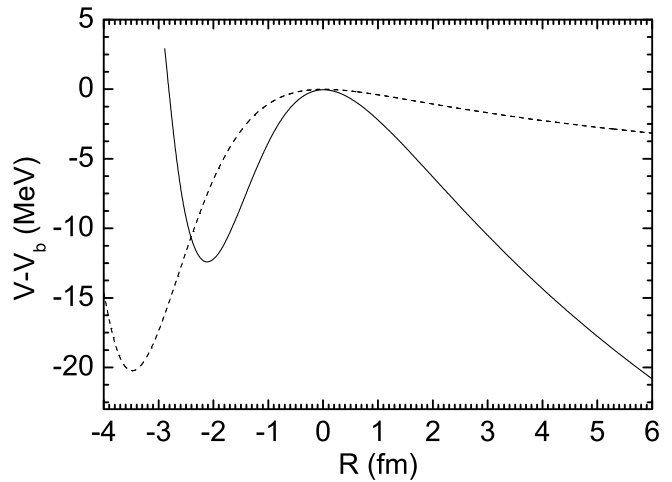}
\caption{The nucleus-nucleus potentials calculated at $J=0$ for the  reactions $^{16}$O+$^{208}$Pb (solid line) and $^{16}$O+$^{16}$O (dashed line).
The coordinate $R$ is defined relative to the position $R_b$ of the Coulomb barrier.}
\label{Fig1}
\end{figure}

The calculated potentials with respect to their  barriers $V_b$ are presented in Fig. \ref{Fig1} for two reactions with
spherical nuclei $^{16}$O+$^{208}$Pb and $^{16}$O+$^{16}$O at $J=0$ .
With increasing angular momentum, the positions  of the potential barrier $R_{b}$
and the minimum $R_m$ merges, and at certain $J$ the potential pocket disappears.
This is a natural limitation of $J$ that contribute to the capture (fusion).
,
The large Coulomb repulsion in the case of $^{16}$O+$^{208}$Pb leads to a steep decline of the potential, compared to that in the case of $^{16}$O+$^{16}$O.
So,   at the fixed $E_{\rm c.m.}-V_b< 0$, two colliding nuclei approach   closer to reach smaller $R_{\rm ext}$ in the case of heavier system.

\section{Extension of the approach}

\subsection{Energy-dependent friction and  internal excitation bandwidth}
\begin{figure}[h]
\centering
\includegraphics[angle=0, width=1\columnwidth]{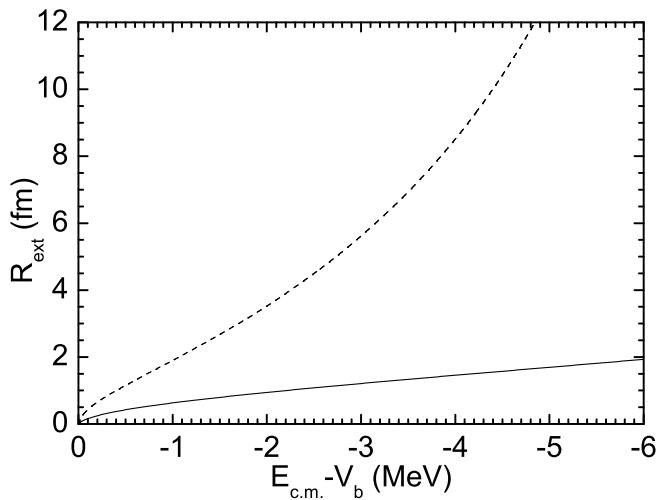}
\caption{The calculated dependencies of the external turning point $R_{\rm ext}$ on $(E_{\rm c.m.}-V_b)$
for the  reactions $^{16}$O+$^{208}$Pb (solid line) and   $^{16}$O+$^{16}$O (dashed line).
The value of $R_{\rm ext}$ is defined relative to the position $R_b$ of the Coulomb barrier.}
\label{Fig2}
\end{figure}

The formalism, introduced in Sect. II, implies that the friction $\lambda$ does not depend on $E_{\rm c.m.}$.
The use of the constant friction seems to be valid in case of   fusion of rather heavy nuclei at energies near and below (up to 5-6 MeV) the Coulomb barrier.
However, in the reactions with medium-mass and light nuclei, and/or at extreme sub-barrier energies, the dependence of the friction
on $E_{\rm c.m.}$ can not be ignored.
This remark can be easily understood from Fig.~\ref{Fig2}, where the comparison of the dependencies of the
external turning point $R_{\rm ext}$ on energy is shown for the reactions $^{16}$O+$^{208}$Pb and $^{16}$O+$^{16}$O.
The  value of  $R_{\rm ext}$ at given $E_{\rm c.m.}$ indicates the degree of the overlap of nuclear density profiles, which is responsible for the nuclear friction.
For the $^{16}$O+$^{16}$O reaction,  the value of $R_{\rm ext}$ drastically increases with decreasing  $E_{\rm c.m.}-V_b$
 which leads to a strong reduction of the friction with respect to the $^{16}$O+$^{208}$Pb reaction.
At fixed  $E_{\rm c.m.}-V_b$, the  value of  $R_{\rm ext}$ is much closer to the position of the
corresponding Coulomb barrier for heavy system.

\begin{figure}[h]
\centering
\includegraphics[angle=0, width=1\columnwidth]{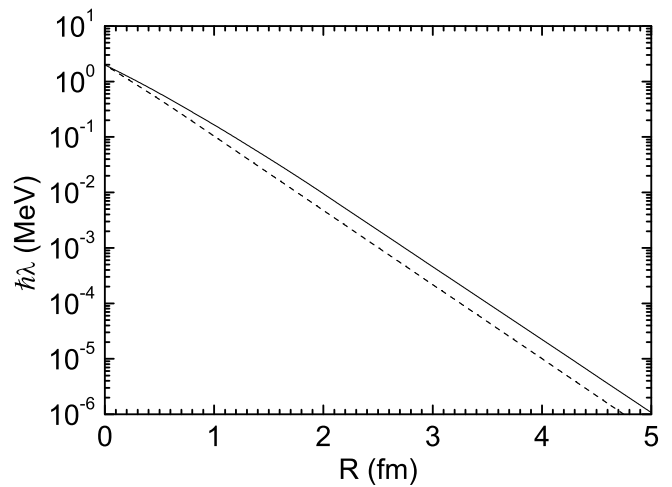}
\caption{The calculated dependencies  of the friction coefficients on $R$ for the  reactions $^{16}$O+$^{208}$Pb (solid line)
and $^{16}$O+$^{16}$O (dashed line).
The coordinate $R$ is defined relative to the position $R_b$ of the Coulomb barrier.}
\label{Fig3}
\end{figure}

To include the bombarding energy dependence of friction in our model, we refer to the studies
of Refs. \cite{GrossKalinovski,weidemuller}, where  the friction,
\begin{eqnarray}
\lambda(R)=\lambda_b\left(\frac{\nabla V_N(R)}{\nabla V_N(R_b)}\right)^2,
\label{lambdaR}
\end{eqnarray}
proportional to the square of  nuclear force, was suggested for fusion and deep inelastic reactions.
This form of $\lambda(R)$ takes into account the overlap of nuclear surfaces on which the friction strength depends.
To determine the normalization parameter $\lambda_b$, we use our previous studies \cite{EPJSub,EPJSub1,EPJSub2,PhysPartNuc2016},
where the fusion cross section of heavy nuclei at energies near and below (up to 4-5 MeV) the Coulomb barrier was well described with
constant friction coefficient $\hbar\lambda_b=\hbar\lambda(R=R_b)=2$ MeV.
The calculated dependencies of the friction on $R$ are shown in
Fig. \ref{Fig3} for the  reactions $^{16}$O+$^{208}$Pb and $^{16}$O+$^{16}$O.
One can see the rapid decrease of the friction with increasing $R$.
Note  that the calculated capture cross sections are rather insensitive to the value of $\lambda_b$.
For example, the variation of this parameter by 2 times leads to the change of the results of the calculations by less then 5$\%$.

In accordance with Eq. (\ref{glo_eq}) the internal excitation bandwidth $\gamma$ is related to the friction.
We take the same relation also in the case of coordinate-dependent friction coefficient $\lambda(R)$:
\begin{eqnarray}
\gamma(R)=\gamma_0\exp\left[k_1\frac{\omega(R)}{\lambda(R)}\right].
\label{gR}
\end{eqnarray}
In the case of constant friction $\hbar\lambda=2$ MeV, the best agreement with the experimental data
is archived at constant internal excitation width $\hbar\gamma=32$ MeV for the reactions with heavy nuclei \cite{EPJSub,EPJSub1,EPJSub2,PhysPartNuc2016}.
Thus, we choose $\gamma_0$ to have  $\hbar\gamma\left(R=R_b\right)=32$ MeV.
Note  that at deep sub-barrier energies the results of calculations are almost unsensitive to $\gamma_0$ (see  subsection IV.D).
 In the limit $\lambda\to 0$, Eq. (\ref{gR}) results in
$\frac{\lambda}{\omega}\ln(\gamma)\to k_1$ as in Eq. (\ref{glo_eq}).

The value of $k_1$ in Eq. (\ref{gR}) is a parameter to be adjusted, and may vary for different reactions.
However, our calculations show a certain universality of this parameter for all considered reactions.
The perfect agreement with the experimental cross sections is archived if the values of $\gamma$, $\omega$, and $\lambda$
are calculated at $R=R_{\rm ext}$
and the value of
$k_1$ is defined as
\begin{eqnarray}
k_1=\frac{\alpha}{\sqrt{\mu\omega^2_b}},
\label{k1}
\end{eqnarray}	
where $\alpha=\frac{\pi }{2}$  MeV$^{1/2}$ fm$^{-1}$
and $\omega_b=\omega(R=R_b)$ is the frequency at the barrier position $R_b$.

So, in our extended model we use the values of friction and internal excitation bandwidth which are calculated at $R=R_{\rm ext}$:
 $\lambda(R_{\rm ext})$ and  $\gamma(R_{\rm ext})$.
Thus, the bombarding energy dependence of $\gamma$ and $\lambda$
 are included through their dependence on $R_{\rm ext}$.

\subsection{Energy-dependent frequency}
We  use the local inverted oscillator approximation
which means that the nucleus-nucleus interaction potential
at each $E_{\rm c.m.}$ is locally replaced by the inverted
oscillator with own frequency. At different $E_{\rm c.m.}$,
there are different local inverted oscillators.
%
%
%
As mentioned in Sect. II,
for the reactions with heavy nuclei at sub-barrier energies,
we determine the frequency $\omega$ of the approximated oscillator from the condition of equality
of the classical actions under the barrier of the real and approximated potentials.
This approximation leads to the close values of $R_{\rm ext}$  for the  real and approximated potentials.
For the reactions with light- and medium-mass nuclei,
the same procedure leads to completely different values of $R_{\rm ext}$ in the cases of real and approximated potentials.
Because the friction strongly depends on $R_{\rm ext}$, this approximation becomes irrelevant.
For the light- and medium-mass nuclei, we suggest to match the height and  position of the barrier of the real potential with
the height and  position of inverted oscillator. To determine the frequency $\omega$ at sub-barrier energies,
we use the following expression
\begin{eqnarray}
V_b-E_{\rm c.m.}=\frac{\mu \omega^2 (R_{\rm ext}-R_b)^2}{2}
\label{E_omega_eq}
\end{eqnarray}	
which provides the dependence
of the frequency $\omega$ on   $E_{\rm c.m.}$  that is on $R_{\rm ext}$.

\subsection{Initial conditions and parameters}
Employing Eq. (\ref{PC_eq}) and the initial coordinate $R_0$ and momentum $P_0$, we
 calculate the capture probability $P_{\rm cap}$.
 Let us consider the initial conditions and parameters
 used in our calculations.

If the collision of nuclei occurs at sub-barrier energies $E_{\rm c.m.}<V_b$, the
dissipation of the kinetic energy of relative motion before $R_{\rm ext}$ is neglected.
Hence, the $R_0$ coincides with the external turning point, $R_0=R_{\rm ext}$, and $P_0=0$.
Here, the values of $\lambda$, $\omega$,  $\gamma$ are calculated at $R=R_0=R_{\rm ext}$,
$\lambda(R_{\rm ext})$,  $\omega(R_{\rm ext})$,  $\gamma(R_{\rm ext})$,
and correspondingly they depend on $E_{\rm c.m.}$.

If the capture occurs at energies $E_{\rm c.m.}$ above the Coulomb barrier $V_b$,
$R_0=R_{b}$ and $P_0=\sqrt{2\mu E_{\rm c.m.}\exp(-2\lambda_b t_{\rm int})}$.
Here, the dissipation $\Delta E=E_{\rm c.m.}[1-\exp(-2\lambda_b t_{\rm int})]$
of the kinetic energy of relative motion
is taken effectively into account by using
the average friction coefficient $\lambda_b$ and  energy-dependent interaction time estimated as
$t_{\rm int}=1/\sqrt{E_{\rm c.m.}}$ s.
For the calculations of $\sigma_{\rm cap}(E_{\rm c.m.})$ at energies above the Coulomb barrier,
we use the values of $\lambda$, $\omega$, and $\gamma$ calculated at the barrier position:
$\hbar\lambda_b=\hbar\lambda(R=R_b)=2$ MeV, $\hbar\gamma_b=\hbar\gamma(R=R_b)=32$ MeV and
$\omega_b=\omega(R=R_b)=\sqrt{\frac{1}{\mu}\frac{d^2V}{dR^2}\mid_{R=R_b}}$.

\subsection{Analytical expression for the capture at extreme sub-barrier energies}
At extreme sub-barrier energies,  we have the following initial conditions:
$P_0=0$ and $R_0=R_{\rm ext}=Z_1Z_2e^2/E_{\rm c.m.}$.
Using this $R_0$ and Eq. (\ref{E_omega_eq}), we obtain the analytical expression
$$\omega=\frac{E_{\rm c.m.}}{Z_1Z_2e^2-R_bE_{\rm c.m.}}\left(\frac{2(V_b-E_{\rm c.m.})}{\mu}\right)^{1/2}$$
for frequency.
Because at extreme sub-barrier energies the value of friction
 is  small and $\gamma\gg\omega$, $\omega/\lambda\gg\ln(\gamma_0)$,  we derive $s_1\simeq\omega$ and
\begin{eqnarray}
\ln\left(\frac{\gamma}{s_1}\right)\simeq\ln(\gamma)\simeq k_1\frac{\omega}{\lambda}.
\label{lngfr}
\end{eqnarray}
Substituting these expressions and  initial conditions  into   Eq. (\ref{PC_eq}), we finally obtain
\begin{eqnarray}
P_{\rm cap}=\frac{1}{2} {\rm erfc}\left[\sqrt\frac{\pi (V_b-E_{\rm c.m.})}{k_1\hbar\omega}\right].
\label{PSFin}
\end{eqnarray}	
Note that Eq. (\ref{PSFin}) is  similar to the well known quantum-mechanical barrier transmission probability
but with the replacement of the usual frequency by the effective one.



\section{Results of calculations}
Using the  procedure  described, we apply  Eqs. (\ref{1a_eq}), (\ref{Asm_fr1})--(\ref{PC_eq}), (\ref{lambdaR}), (\ref{gR}), and (\ref{E_omega_eq})
to calculate the capture cross-section $\sigma_{\rm cap}(E_{\rm c.m.})$ for low-energy
reactions with   light- and medium-mass nuclei.
As emphasized in
\cite{JiangC12C12,HighC12C12,MazarakisC12C12,AguileraC12C12,ChristensenC12O16,ThomasO16O16,HulkeO16O16}, the fusion reactions
between carbon and oxygen isotopes are playing a crucial role in a wide variety of stellar burning scenarios.
As the first step in that direction, we compare  our calculated results with the available data.

\begin{figure}[h]
\centering
\includegraphics[angle=0, width=1\columnwidth]{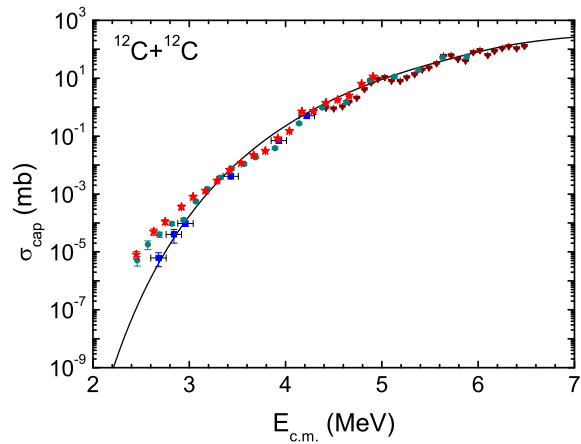}
\caption{ The calculated capture cross section (line) vs $E_{\rm c.m.}$ for the  $^{12}$C+$^{12}$C reaction
compared with the available experimental data.  The experimental data marked by squares, circles, stars and triangles are taken from
Refs.~\protect\cite{JiangC12C12,HighC12C12,MazarakisC12C12,AguileraC12C12}, respectively}
\label{FigC12C12}
\end{figure}

\begin{figure}[h]
\centering
\includegraphics[angle=0, width=1\columnwidth]{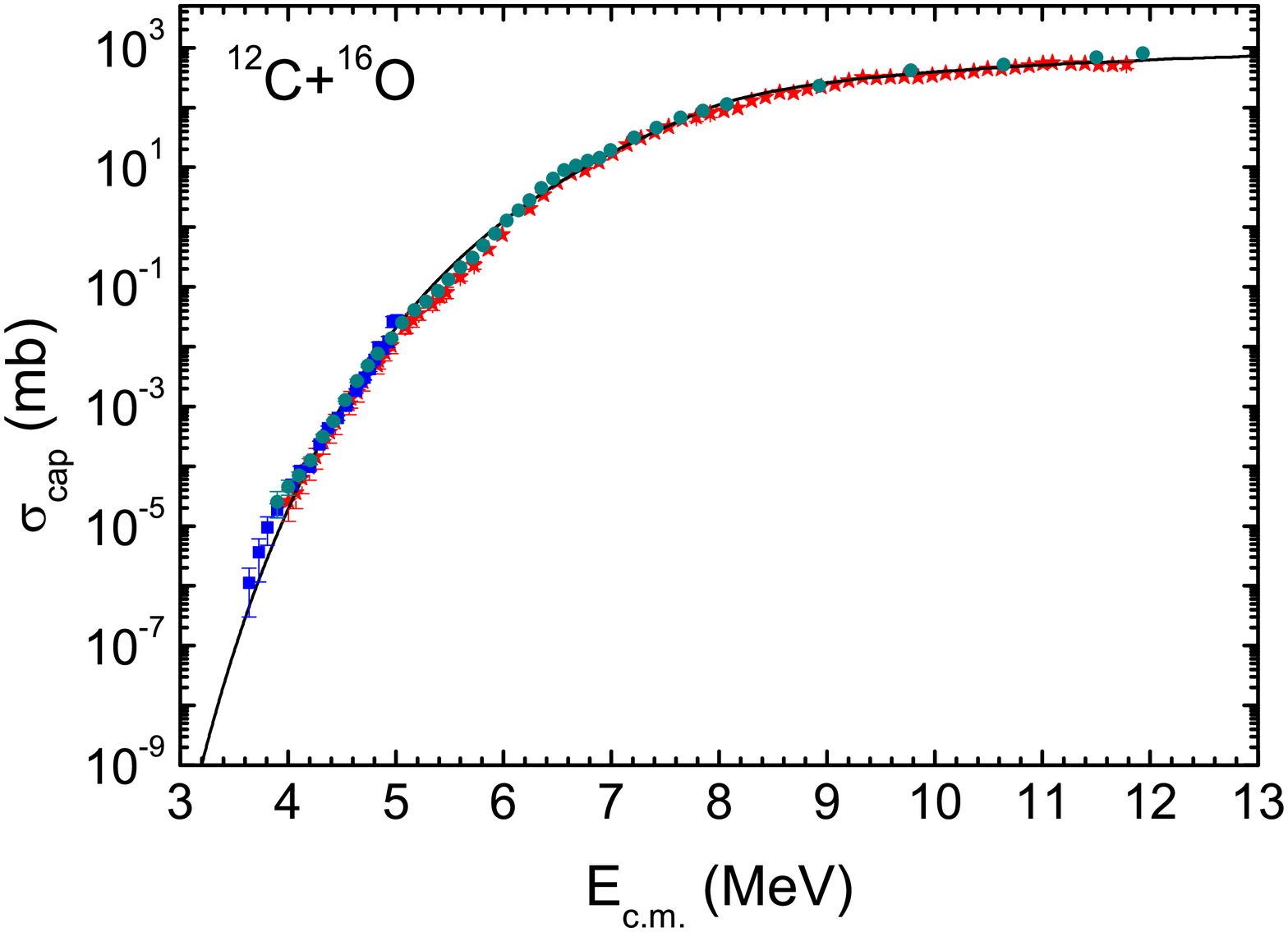}
\caption{ The same as in Fig. \ref{FigC12C12}, but for the $^{12}$C+$^{16}$O reaction.
The experimental data marked by squares, circles, and stars are taken from
Refs.~\protect\cite{FangC12O16,ChristensenC12O16,CujecC12O16}, respectively.}
\label{FigC12O16}
\end{figure}

\begin{figure}[h]
\centering
\includegraphics[angle=0, width=1\columnwidth]{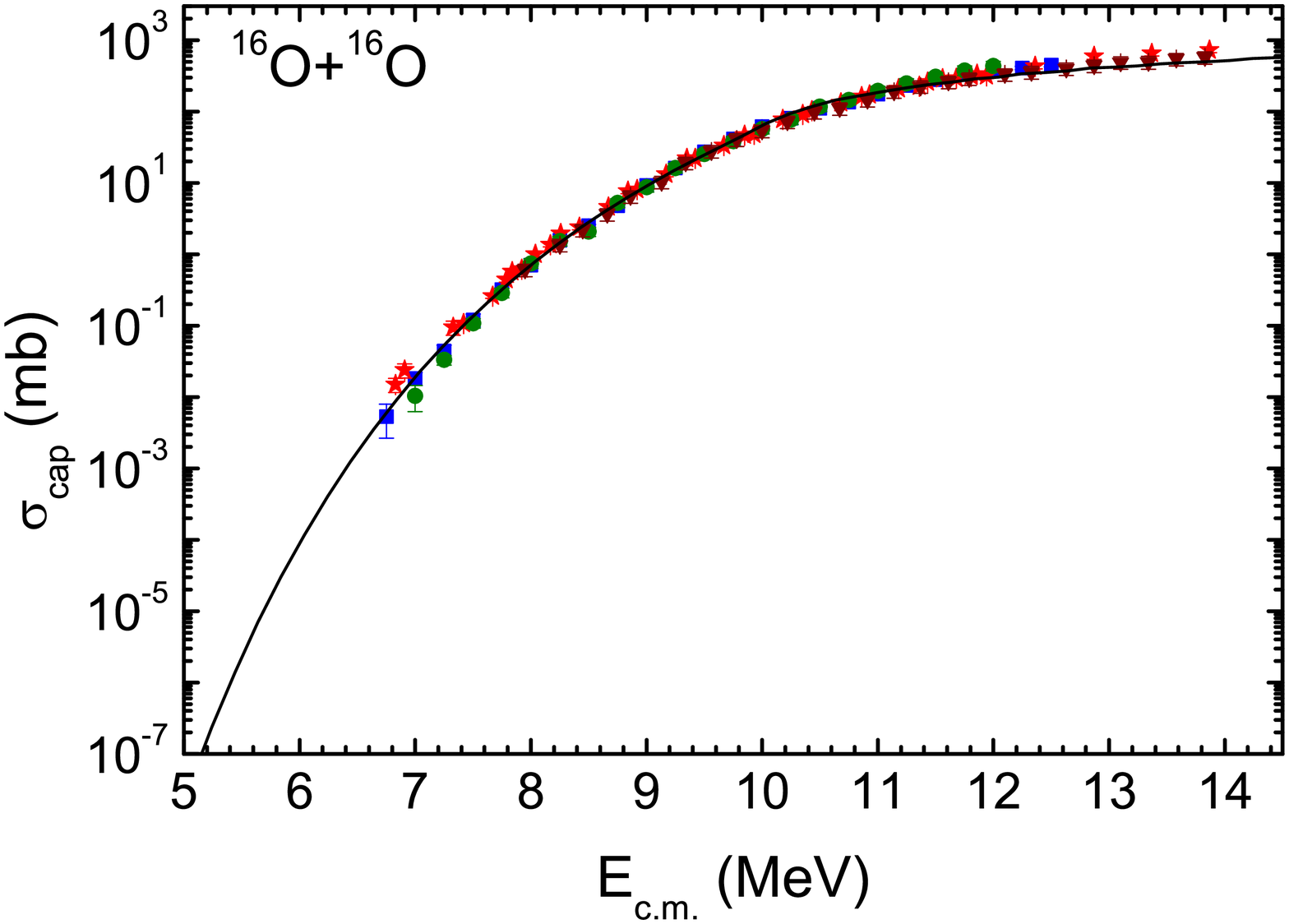}
\caption{ The same as in Fig. \ref{FigC12C12}, but for the  $^{16}$O+$^{16}$O reaction.
The experimental data marked by squares, circles, triangles and stars are taken from
Refs.~\protect\cite{ThomasO16O16,WuO16O16,KuronenO16O16,HulkeO16O16}, respectively.}
\label{FigO16O16}
\end{figure}

\begin{figure}[h]
\centering
\includegraphics[angle=0, width=1\columnwidth]{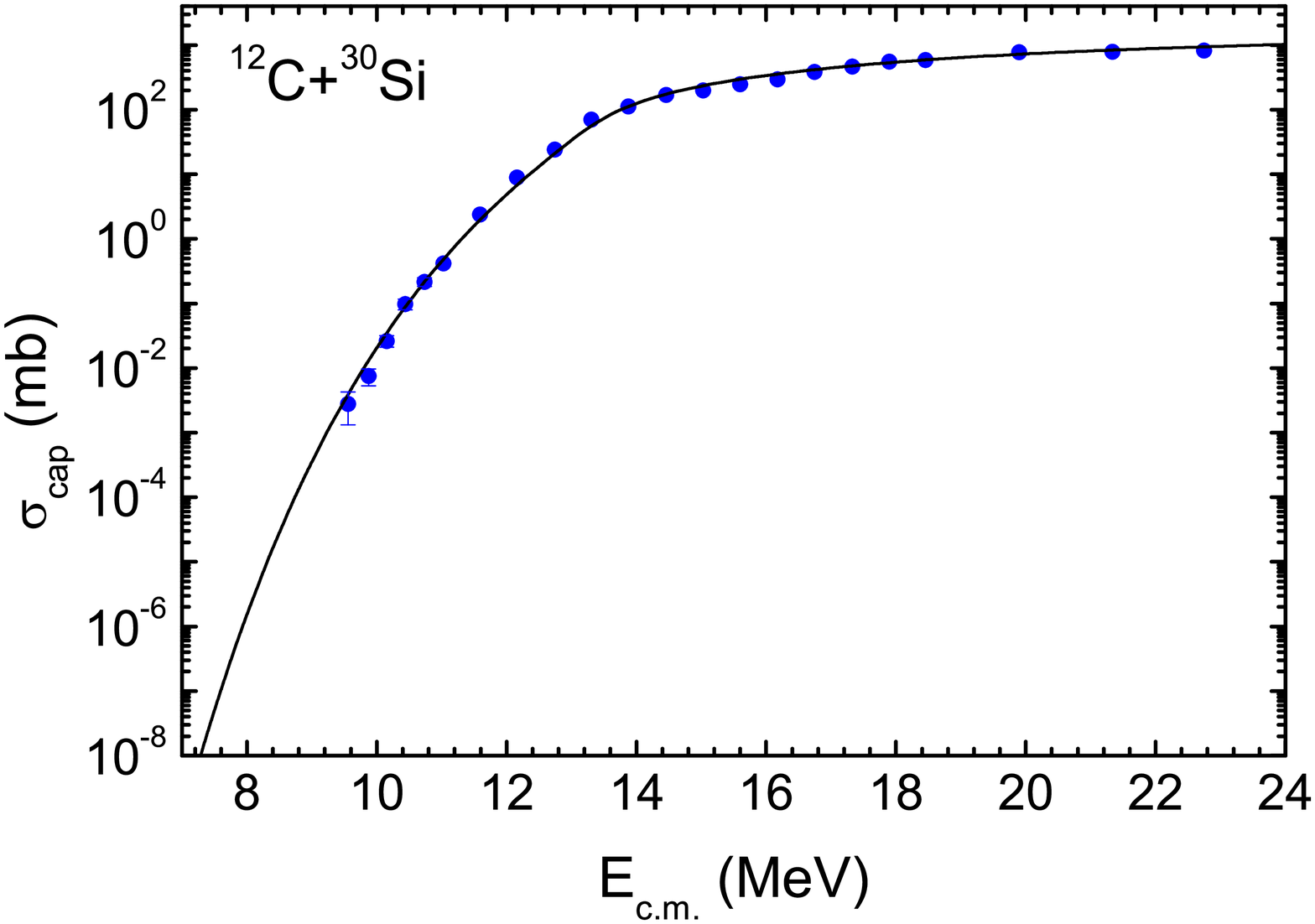}
\caption{  The same as in Fig. \ref{FigC12C12}, but for the $^{12}$C+$^{30}$Si reaction.
The experimental data marked by circles are taken from Ref.~\protect\cite{MontagnoliC12Si30}.}
\label{FigC12Si30}
\end{figure}

\begin{figure}[h]
\centering
\includegraphics[angle=0, width=1\columnwidth]{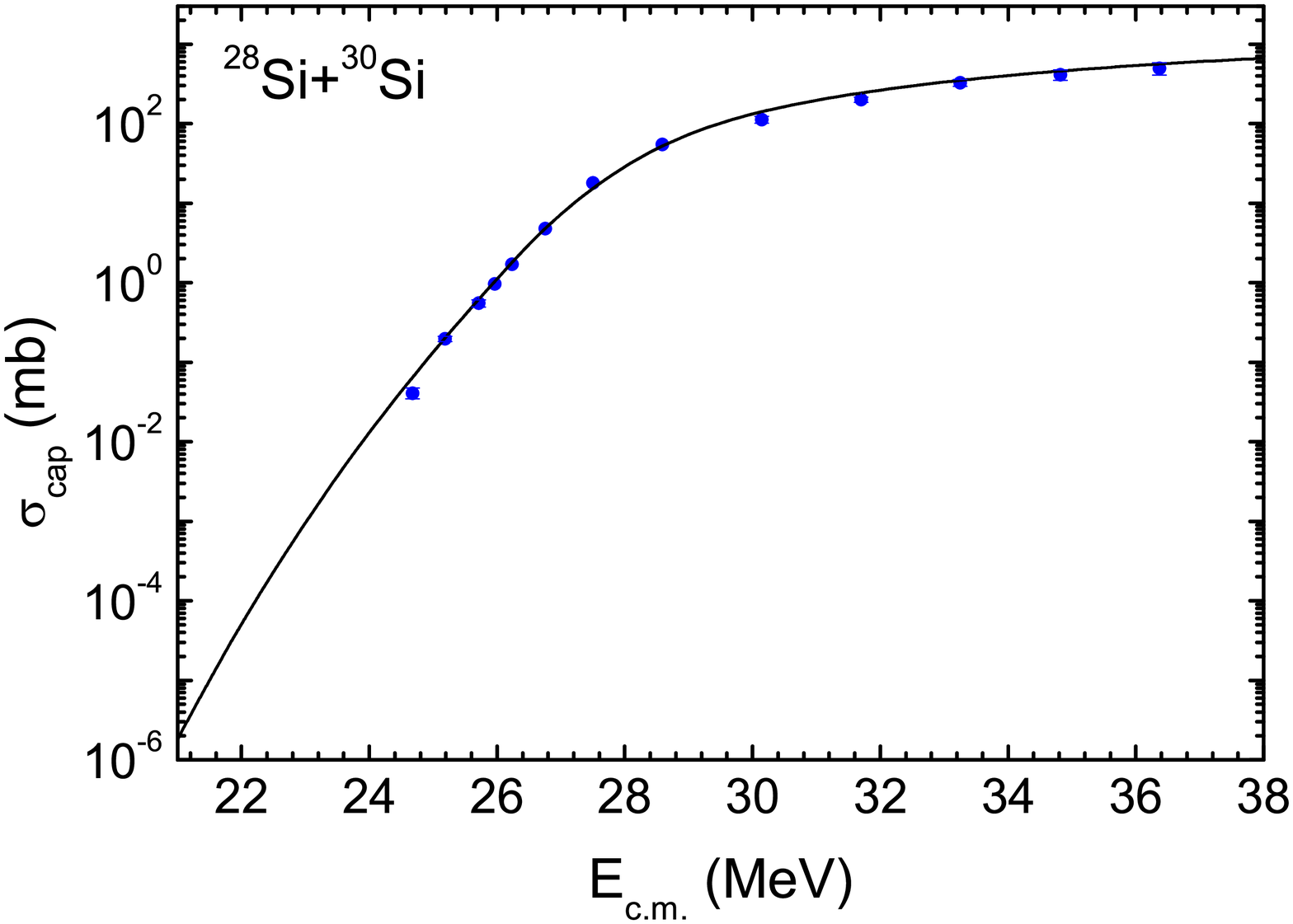}
\caption{  The same as in Fig. \ref{FigC12C12}, but for the $^{28}$Si+$^{30}$Si reaction.
The experimental data marked by circles are taken from Ref.~\protect\cite{JiangSi28Si30}.}
\label{FigSi28Si30}
\end{figure}

\begin{figure}[h]
\centering
\includegraphics[angle=0, width=1\columnwidth]{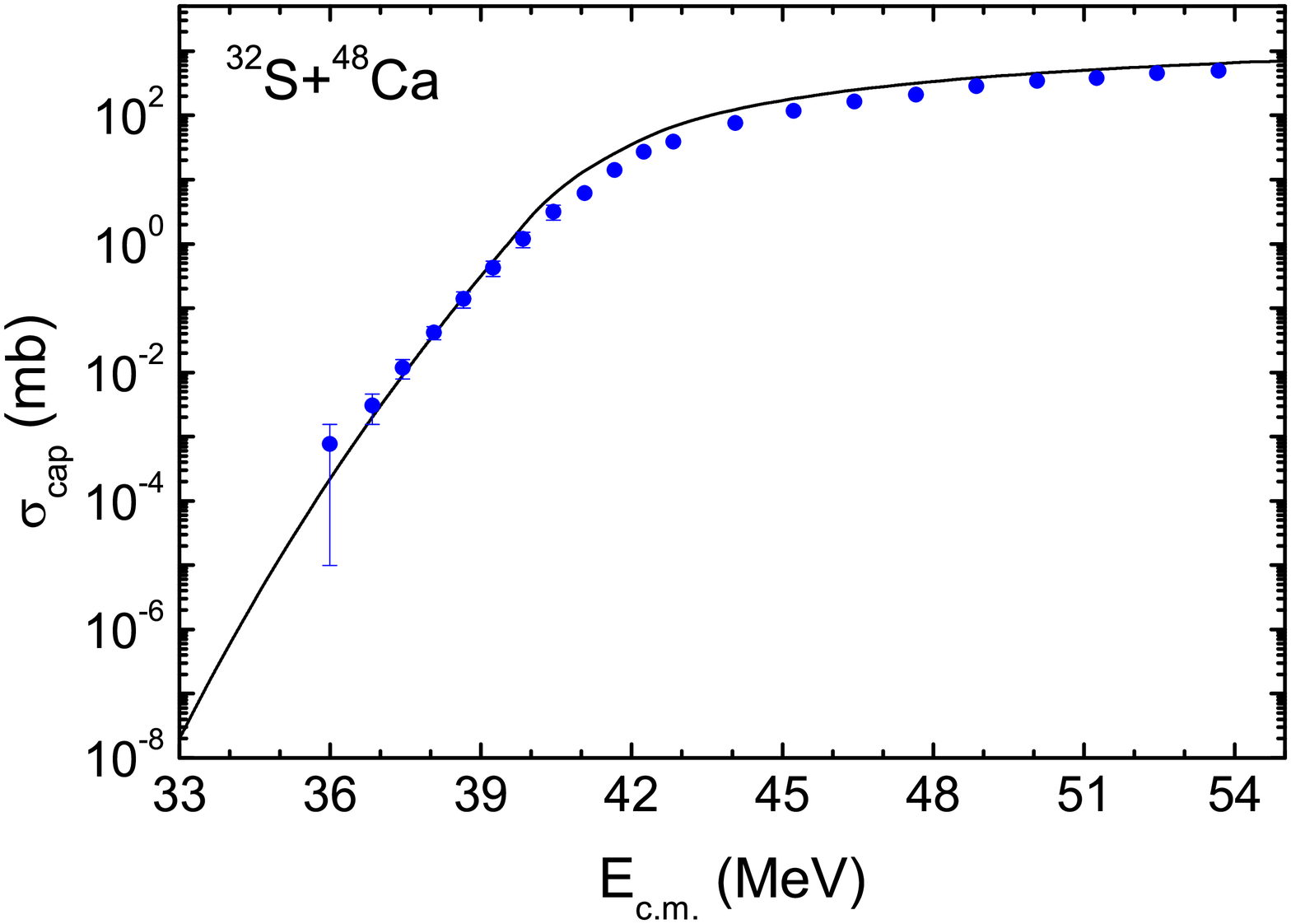}
\caption{  The same as in Fig. \ref{FigC12C12}, but for the $^{32}$S+$^{48}$Ca reaction.
The experimental data marked by circles are taken from Ref.~\protect\cite{MontagnoliS32Ca48}.}
\label{FigS32Ca48}
\end{figure}

\begin{figure}[h]
\centering
\includegraphics[angle=0, width=1\columnwidth]{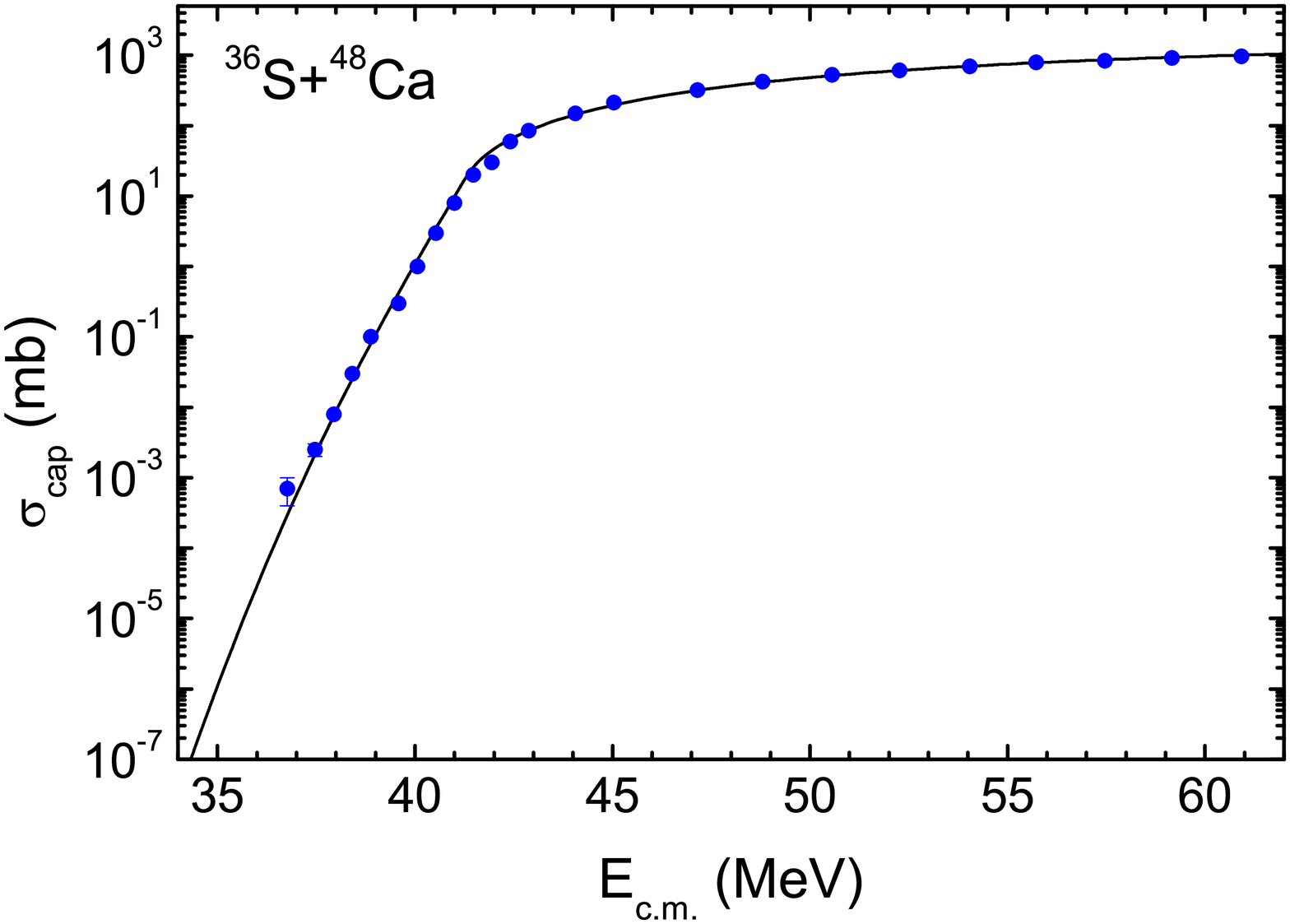}
\caption{  The same as in Fig. \ref{FigC12C12}, but for the $^{36}$S+$^{48}$Ca reaction.
The experimental data marked by circles are taken from Ref.~\protect\cite{StefaniniS36Ca48}.}
\label{FigS36Ca48}
\end{figure}

The results of the calculated capture cross sections and the experimental data are shown in Figs. 4--10.
In all considered reactions we obtain a good agreement with the experiments.
Note, that for  $^{12}$C+$^{12}$C reaction the early measured data \cite{HighC12C12,MazarakisC12C12}
differ from the later ones \cite{JiangC12C12,AguileraC12C12}. Here,
the mechanism that causes the oscillations of the cross section in the   $^{12}$C+$^{12}$C
reaction is not considered \cite{AlexisC12C12}.

Our calculated results at sub-barrier energies are rather sensitive  to the coefficient $k_1$ [Eq.(\ref{k1})].
However, it is uniformly determined for all reactions considered. Thus, we conclude that Eq. (\ref{k1}) is useful
for the reactions of astrophysical interest.

\begin{figure}
\includegraphics[angle=0, width=1\columnwidth]{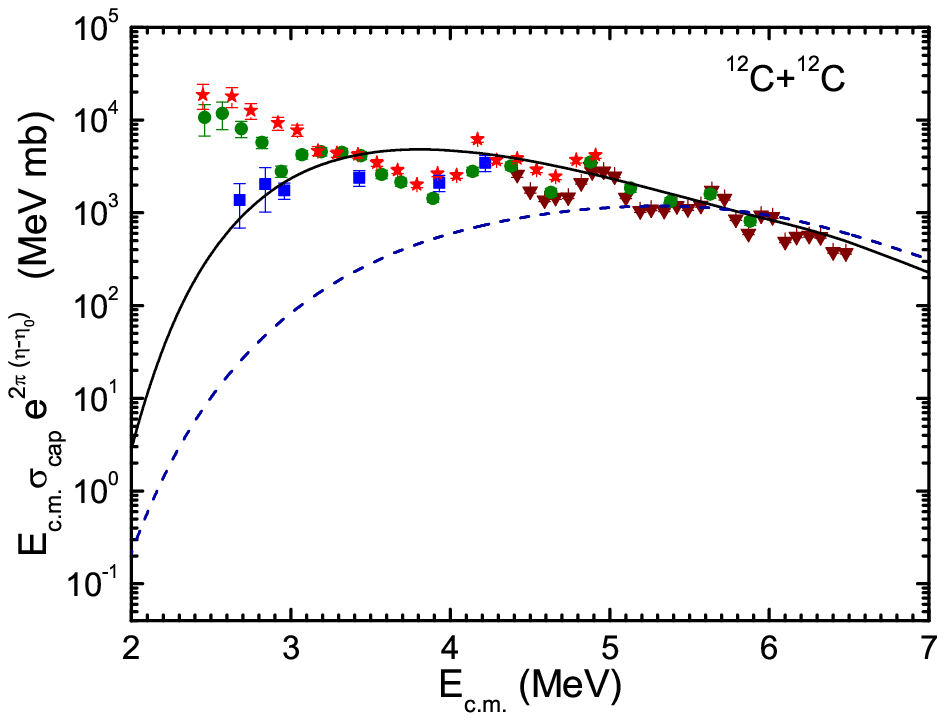}
\includegraphics[angle=0, width=1\columnwidth]{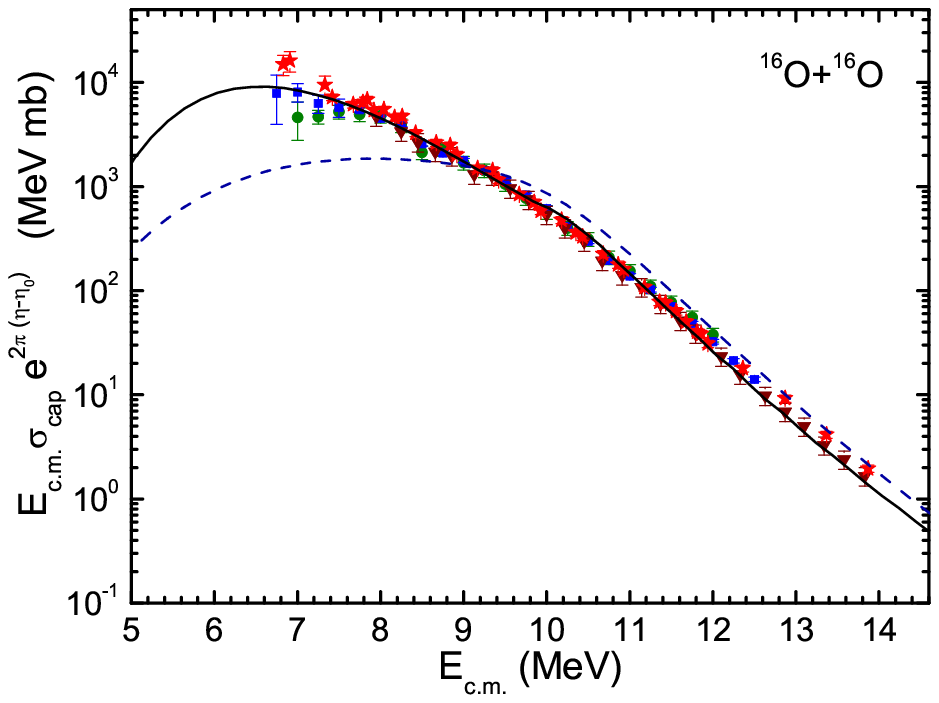}
\caption{The calculated   astrophysical $S$-factor   vs $E_{\rm c.m.}$
for the  reactions $^{12}$C+$^{12}$C ($\eta_0=\eta(E_{\rm c.m.}=V_b)=5.58$) and $^{16}$O+$^{16}$O ($\eta_0=9.01$) (solid lines).
Comparison of $S$-factors from the WKB model (dashed lines).
The experimental data (symbols)
are from Refs.~\protect\cite{JiangC12C12,HighC12C12,MazarakisC12C12,AguileraC12C12,ThomasO16O16,WuO16O16,KuronenO16O16,HulkeO16O16}.
}
\label{6_fig}
\end{figure}

\begin{figure}
\includegraphics[angle=0, width=1\columnwidth]{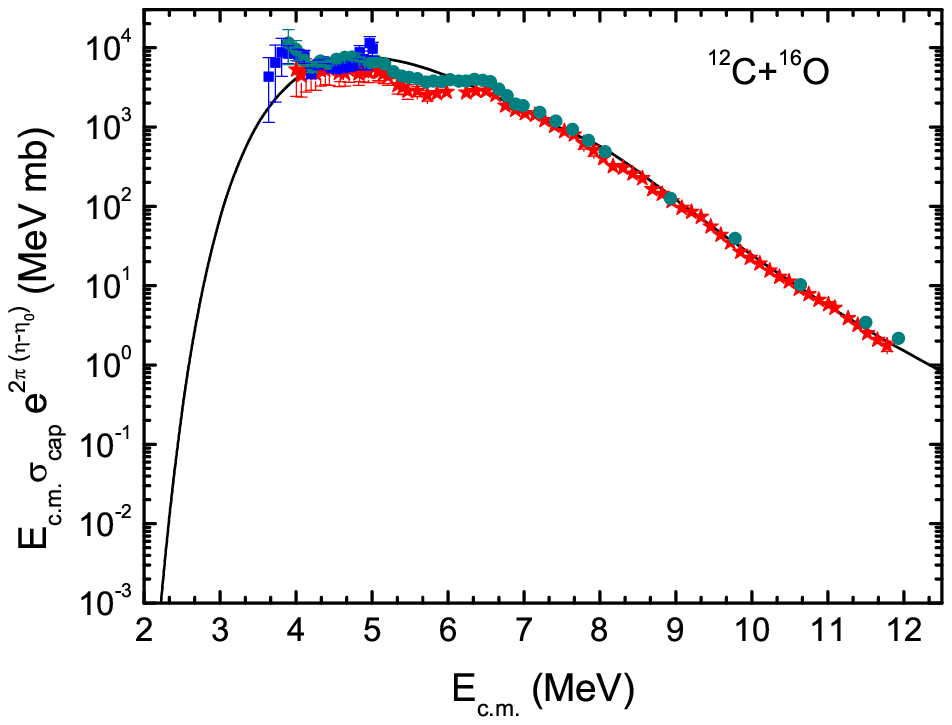}
\includegraphics[angle=0, width=1\columnwidth]{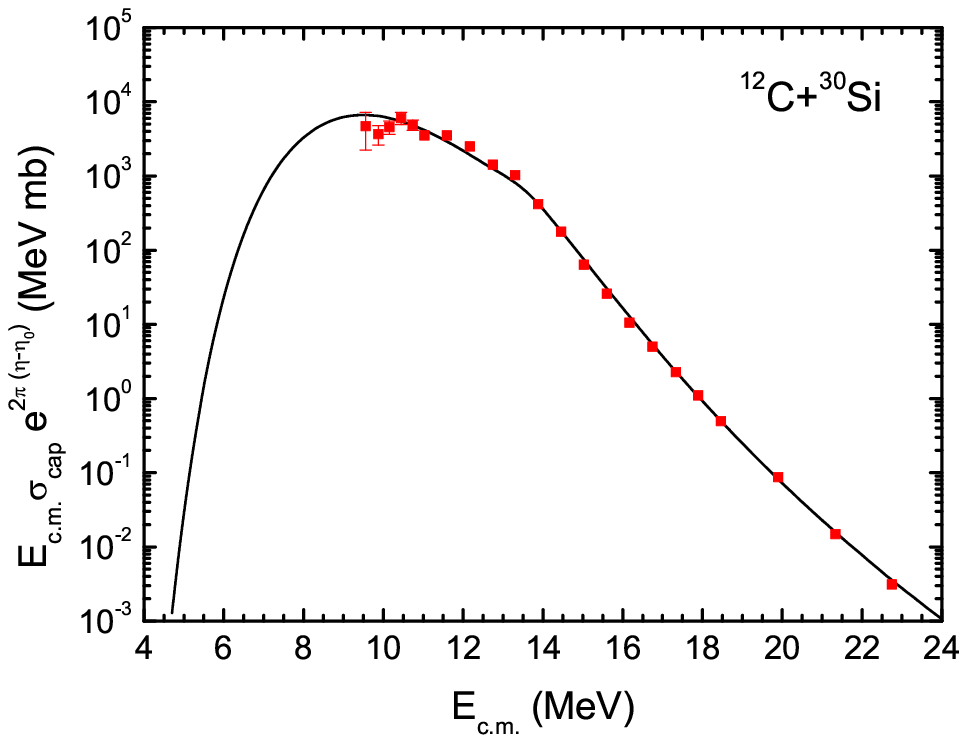}
\includegraphics[angle=0, width=1\columnwidth]{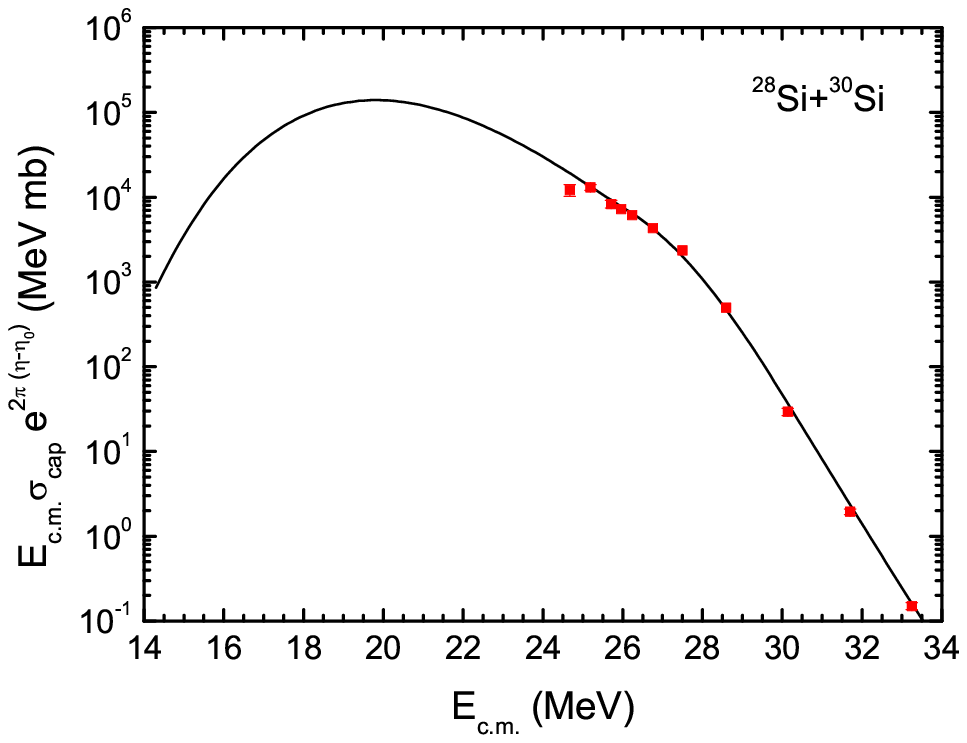}
\caption{The calculated astrophysical $S$-factor  vs $E_{\rm c.m.}$ for the  reactions
$^{12}$C+$^{16}$O ($\eta_0=7.07$),   $^{12}$C+$^{30}$Si ($\eta_0=10.60$),
and $^{28}$Si+$^{30}$Si ($\eta_0=22.15$).  Here, the calculated values are normalized
to the experimental data~\protect\cite{FangC12O16,ChristensenC12O16,CujecC12O16,MontagnoliC12Si30,JiangSi28Si30}.
}
\label{7_fig}
\end{figure}

At energies below the Coulomb barrier, where the cross section drops rapidly with decreasing energy, it is more convenient
to discuss the astrophysical $S$-factor,
\begin{eqnarray}
S(E_{\rm c.m.})=E_{\rm c.m.}\sigma_{\rm fus}(E_{\rm c.m.})\exp[2\pi(\eta-\eta_0)],
\label{Sfactor}
\end{eqnarray}
rather than the fusion excitation function.
Here, $\eta(E_{\rm c.m.})=Z_1 Z_2 e^2\sqrt{\mu/(2\hbar^2E_{\rm c.m.})}$ is the Sommerfeld parameter and
$\eta_0=\eta(E_{\rm c.m.}=V_b)$,  where $V_b$ is the Coulomb barrier height for the spherical
interacting nuclei.
Assuming that the capture cross section is equal to the fusion
cross section, we calculate the astrophysical $S$-factor.
In Figs. 11 and 12  the calculated  $S$-factors versus $E_{\rm c.m.}$ are shown
for the reactions $^{12}$C+$^{12}$C, $^{12}$C+$^{16}$O,   $^{12}$C+$^{30}$Si, $^{16}$O+$^{16}$O,
and $^{28}$Si+$^{30}$Si. A good agreement of the calculated  excitation function with
the experimental data leads to a good description of $S$-factor as well. For the reactions under study,
the $S$-factor has a maximum at $E_{\rm c.m.}\approx\frac{2}{3}V_b$,
where $V_b$ is the Coulomb barrier height for the spherical interacting nuclei.
The origin of the maximum of the $S$-factor is the turning-off of the nuclear forces between the colliding nuclei with decreasing $E_{\rm c.m.}$.
While the theory shows clear maximum, their presence in the
experimental data is tenuous up to now.
In the recent paper \cite{JiangC12C12} on a new measurement of the $^{12}$C+$^{12}$C fusion cross sections,
it was found that the astrophysical S-factor exhibits a maximum around $E_{\rm c.m.}$=3.5--4 MeV.
The additional measurements of different systems at lowest  bombarding energies are necessary
to establish the existence of $S$-factor maximum.
In Figs. 11 and 12, after this maximum $S$-factor   decreases strongly with decreasing bombarding energy, which leads to a reduction of
the previously predicted astrophysical reaction rates.
Note also that such a strong dependence on $E_{\rm c.m.}$, in fact, contradicts the philosophy of representing the cross section through the $S$-factor.

Figure 11 shows a comparison between our and WKB ($P_{\rm cap}$ is determined within both the WKB model and the interaction potential of Eqs. (\ref{pot})--(\ref{32ab_eq}))
$S$-factors for
the reactions $^{12}$C+$^{12}$C and $^{16}$O+$^{16}$O.
As seen, the fluctuation and dissipation effects taken into account in our model increase fusion (capture) probability
at sub-barrier energies and decrease at above barrier energies.

\section{Summary}

In the collisions of light- and medium-mass nuclei at low sub-barrier energies, the external turning point is located far from the Coulomb barrier position.
This means a weak overlap of nuclear surfaces and, correspondingly, small   friction.
To this end, we extended our quantum diffusion approach and considered the friction  depending on the bombarding energy.
Using the extended approach, we compared the calculated capture cross-sections with the available experimental data.
In all cases we obtained a good description of the experiments. Comparing the fusion excitation functions calculated within the quantum diffusion
and WKB approaches, we found that the  the fluctuation and dissipation increase fusion  cross section
at sub-barrier energies.
For the reactions $^{12}$C+$^{12}$C, $^{12}$C+$^{16}$O,   $^{12}$C+$^{30}$Si, $^{16}$O+$^{16}$O,
and $^{28}$Si+$^{30}$Si, the maximum of astrophysical $S$-factor at $E_{\rm c.m.}\approx\frac{2}{3}V_b$
was predicted. However, more experimental data at low energies is needed to confirm our predictions.
Another interesting behavior of the obtained S-factor is that its dependence on $E_{\rm c.m.}$
is quite strong at the collision energies below the maximum.

In the limit of weak friction, which corresponds to extreme sub-barrier energies, the analytic expression (\ref{PSFin}) for the
capture probability is obtained. This simple expression can be applied to the reactions of astrophysical interest.
It determines the reaction rates from which, in turn, the astrophysical $S$-factors are derived.
The strong decline of fusion cross sections at sub-barrier energies  considerably reduces
the stellar burning rates and, moreover, leads to severe experimental problems, inhibiting the measurements in many cases.
 This demands for the reliable theoretical methods, allowing us to extrapolate $\sigma_{\rm cap}(E_{\rm c.m.})$ into the experimentally
inaccessible regions at extreme
sub-barrier energies.\\

V.V.S. acknowledges the Alexander von Humboldt-–Stiftung (Bonn).
This work was partially supported by Russian Foundation for Basic Research
(Moscow, grant number 17-52-12015) and DFG (Bonn, contract Le439/16).


\end{document}